# Evolving Antennas for Ultra-High Energy Neutrino Detection

**Julie A. Rolla*** **on behalf of the GENETIS Collaboration**
(a complete list of authors can be found at the end of the proceedings)

*The Ohio State University,*
*281 W Lane Ave, Columbus, OH, 43210*
*United States of America*

*E-mail:* rolla.3@osu.edu

Evolutionary algorithms are a type of artificial intelligence that utilize principles of evolution to efficiently determine solutions to defined problems. These algorithms are particularly powerful at finding solutions that are too complex to solve with traditional techniques and at improving solutions found with simplified methods. The GENETIS collaboration is developing genetic algorithms (GAs) to design antennas that are more sensitive to ultra-high energy neutrino-induced radio pulses than current detectors. Improving antenna sensitivity is critical because UHE neutrinos are rare and require massive detector volumes with stations dispersed over hundreds of km$^2$. The GENETIS algorithm evolves antenna designs using simulated neutrino sensitivity as a measure of fitness by integrating with XFdtd, a finite-difference time-domain modeling program, and with simulations of neutrino experiments. The best antennas will then be deployed in-ice for initial testing. The GA's aim is to create antennas that improve on the designs used in the existing ARA experiment by more than a factor of 2 in neutrino sensitivities. This research could improve antenna sensitivities in future experiments and thus accelerate the discovery of UHE neutrinos. This is the first time that antennas have been designed using GAs with a fitness score based on a physics outcome, which will motivate the continued use of GA-designed instrumentation in astrophysics and beyond. This proceeding will report on advancements to the algorithm, steps taken to improve the GA performance, the latest results from our evolutions, and the manufacturing road map.

*Presenter

## 1. Introduction

Neutrino astronomy is a field of particle astrophysics that seeks to investigate distant phenomena through the detection of neutrinos. As weakly interacting particles, neutrinos are able to travel cosmic distances directly from their source, thus providing a unique means of exploring particle sources in the distant universe [1, 2]. The field of ultra-high energy (UHE) neutrino astronomy probes the most energetic events at above $10^{18}$ eV. Due to the low neutrino cross-section, UHE neutrino detection experiments rely on massive detector volumes, such as the Antarctic ice, to detect when a neutrino directly strikes an atom. The collision of a neutrino within the ice produces Askaryan radiation. Askaryan radiation is of particular interest, as it produces radio signals with attenuation lengths in pure ice on the order of 1 km [3].

The GENETIS (Genetically Evolving NEuTrIno TeleScopes) project focuses on using genetic algorithms (GAs) to improve the sensitivities of Askaryan radiation-based detectors, such as those used by ANITA, ARA, and ARIANNA. These experiments utilize various types of antennas to detect the neutrino-induced radio impulses [2, 4, 5]. GAs are computational techniques that mimic the mechanisms of biological evolution by creating populations of potential solutions to a defined problem and then evolving the population over multiple generations to find optimized solutions [6, 7]. Our research seeks to improve experimental sensitivities with consideration to current constraints in the geometry of antennas deployed in the ice, and the signal characteristics of neutrino-generated Askaryan radiation.

These proceedings describe some background information regarding GAs, as well as two main projects: (1) the Physical Antenna Evolution Algorithm (PAEA) and (2) the Antenna Response Evolution Algorithm (AREA). In PAEA, the physical properties of bicone antennas are evolved for improved neutrino sensitivities. AREA is focused on evolving only gain patterns that are better suited for neutrino detection.

## 2. Genetic Algorithms

GAs are a type of evolutionary computation used to discover one or more sets of values that provide a best-fit to a large parameter space that would have otherwise been difficult or not possible through more traditional techniques [6, 8–10]. Each generation consists of a population of solutions, each of which is tested for its output against a predefined set of goals. To quantify the performance of an individual, a fitness function is created to generate a fitness score to test an individual against the optimal or desired goals. Each generation of individuals has the potential, but is not guaranteed, to improve upon the previous generation [11].

The GA workflow is presented in Figure 1. First, a population of individuals is generated, each with randomly generated genes. Each individual is tested through the fitness function to generate a fitness score. A selection method is implemented to decide which individuals, called parents, will be used in creating the next generation. An operator then uses the parents' genetic code to create offspring individuals for the next generation. This process is iterated until a specified fitness score threshold is surpassed by one or more individuals, or until a specified number of generations have passed.

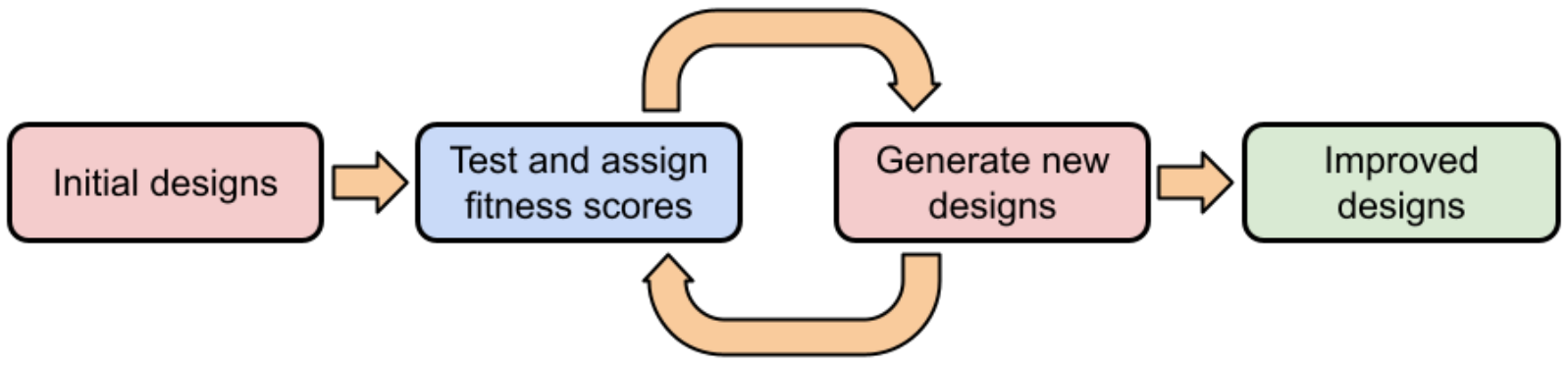

**Figure 1:** General GA evolution procedure.

GAs often combine different selection methods and operators. The two most common selection methods are roulette and tournament selection [12]. In roulette selection, the probability of an individual being selected is proportional to the individual's fitness score. In tournament selection, individuals are randomly placed into groups (tournaments) and compared by fitness score. The individual(s) with the highest fitness score in each tournament are then selected. Additionally, three primary operators are used in this research: crossover, mutation, and reproduction. In crossover, individuals swap genes to form the offspring. In mutation, genetic diversity is introduced altering genes [13]. Finally, in reproduction an individual is passed directly to the next generation without modification.

### 2.1 Genetic Algorithm Parameter Optimization

The proportions of selection methods and genetic operators used can have a large effect on the performance of a GA — typically measured by time to converge to a solution and the maximum fitness score achieved. To optimize these parameters for the GA used in Section 3, a simplified framework was created to assess the performance of the GA with different parameters.

The simplified structure evolved antennas to a specific design. The fitness score was a measure of how similar a generated antenna was to the specified goal. This was done by averaging the magnitude of the differences between each of a generated antenna's genes and the desired design's genes:

$$F = 100 - \sum_{i=1}^{N} \frac{|G_i - g_i|}{N}.$$

Here, the individual's fitness score $F$ is equal to 100 when a generated antenna perfectly matches the goal design. $G_i$ is the value of the i$^{th}$ gene of the goal antenna, $g_i$ is the value of the i$^{th}$ gene of the generated antenna, and $N$ is the total number of genes.

This investigation tested various combinations of GA parameters and found that the ideal combination of selection methods is 80% roulette and 20% tournament. For the genetic operators, the optimal combination is 70% crossover, 24% mutation, and 6% reproduction.

## 3. Physical Antenna Evolution Algorithm (PAEA)

The goal of this investigation is to evolve physical antennas for the best sensitivity to neutrinos in an in-ice array. PAEA evolves antenna geometries by means of fitness scores based on the simulation of antenna performance. For PAEA, a seed generation of possible antenna designs is initialized, with values for each gene picked from uniform distributions, as presented in Table 1. The ranges of the uniform distributions are determined by the constraints of the antenna design.

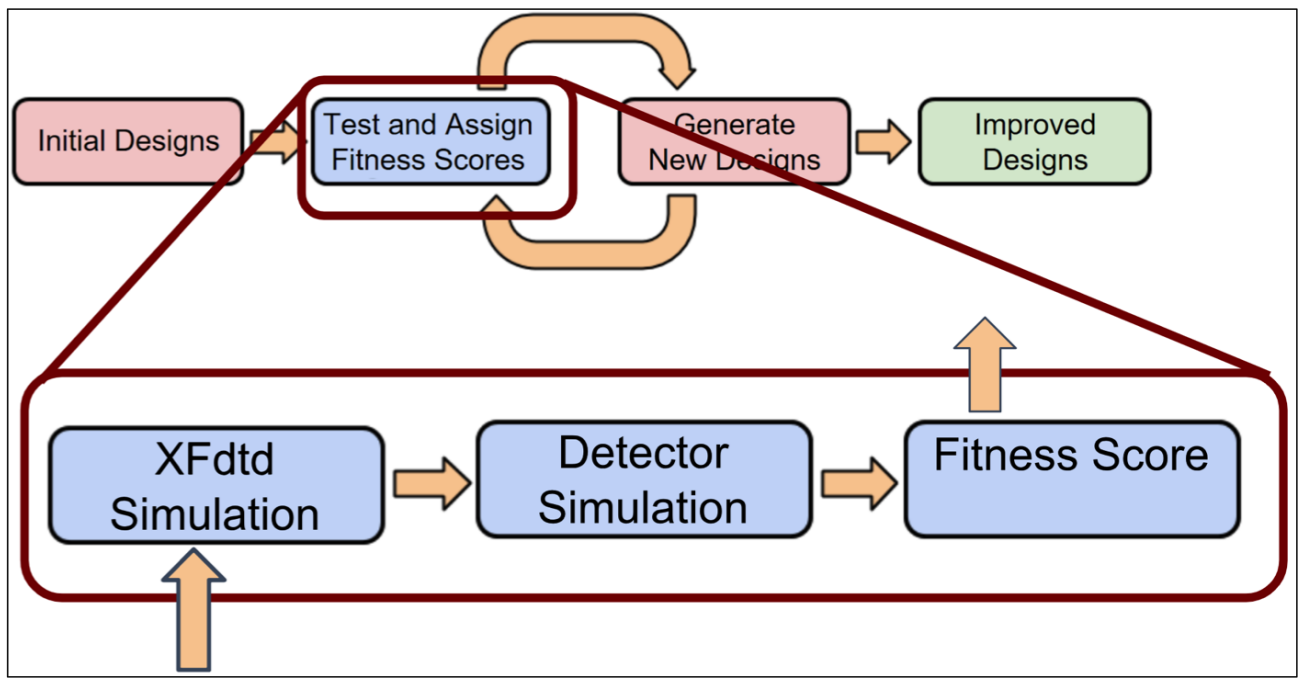

**Figure 2:** A flow chart of the PAEA procedure with a detailed inset of the fitness score procedure.

The commercial antenna simulation software, XFdtd, is then used to model the associated frequency-dependent response patterns. Those antenna response patterns are input into Monte Carlo neutrino simulation software. AraSim, in order to predict the effective volume of ice the detector is sensitive to when that antenna is used [14]. The fitness score for each antenna is fed back into the GA that evolves the geometry of the next generation of antennas. The PAEA process is generic and can be adapted for many types of antenna designs.

In the current version of PAEA, we consider an asymmetric bicone antenna design as shown in Figure 3. The bicone design consists of two truncated cones sitting back-to-back with a fixed separation distance of 3.0 cm. The parameters of the design that are subject to evolution are: the minor radius of the cone, the length of the cone, and the cone's opening angle. Each cone can have different values for each parameter, for a total of six values that define the antenna.

**Table 1:** Summary of the mean and standard deviation for the newly selected parameters drawn from a Gaussian distribution

| Gene | $\mu$ | $\sigma$ |
|---|---|---|
| Length | 89.92 | 78.36 |
| Minor Radius | 2.08 | 6.12 |
| Opening Angle | 0.0140 | 0.0017 |

In the results presented, the GENETIS algorithm evolved 50 individuals over 15 generations. The results of the algorithm are presented in the violin plot in Figure 4, which shows evolution toward improved solutions. For each generation, the range from best and worst fitness scores are illustrated by the vertical lines, with the lowest individual being at the bottom of the line and the highest at the top. The vertical lines are not error bars for the mean fitness score. The width of the line represents the probability density of the population. The current fitness of the Ara antenna is shown as the black dotted line for reference. The antenna with the highest fitness thus far is shown in Figure 5.

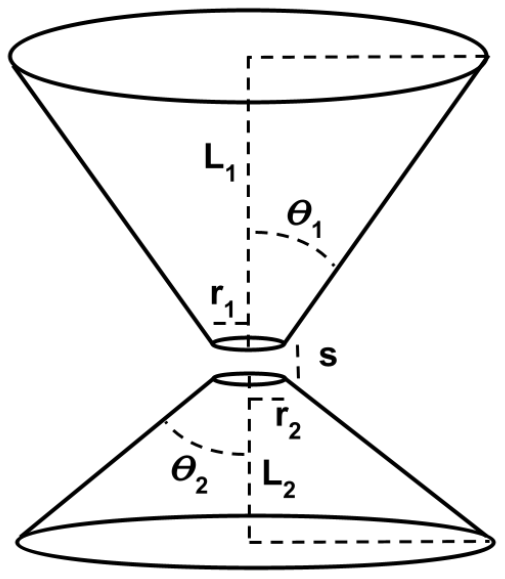

**Figure 3:** The geometry of an asymmetric bicone antenna. The lengths ($L_1$, $L_2$), minor radii ($r_1$, $r_2$), opening angles ($\theta_1$, $\theta_2$), and separation distance ($s$) fully define the geometry. In the results presented here, the separation distance was held constant, and the other six parameters were varied.

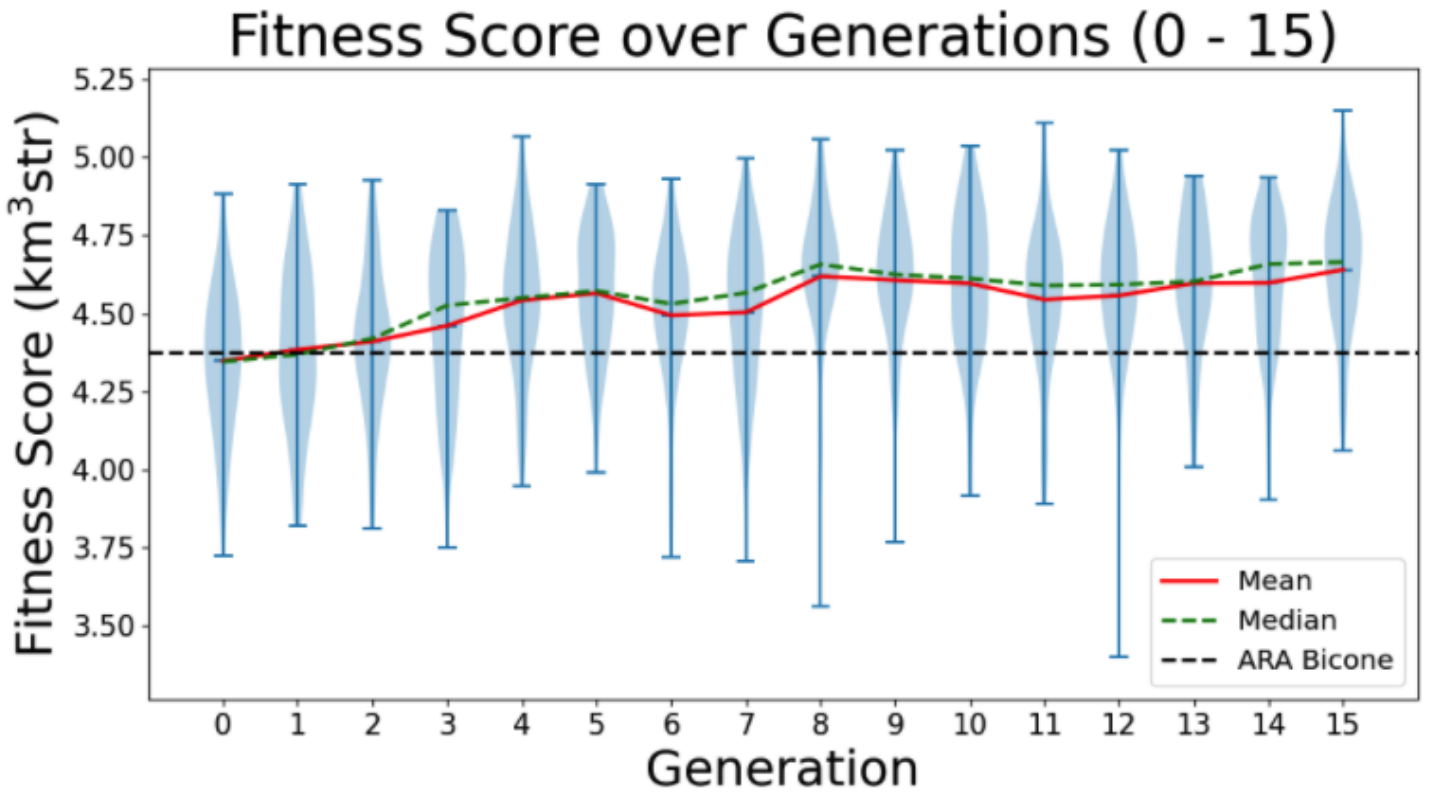

**Figure 4:** Initial results from PAEA showing evolution to antennas with improved fitness.

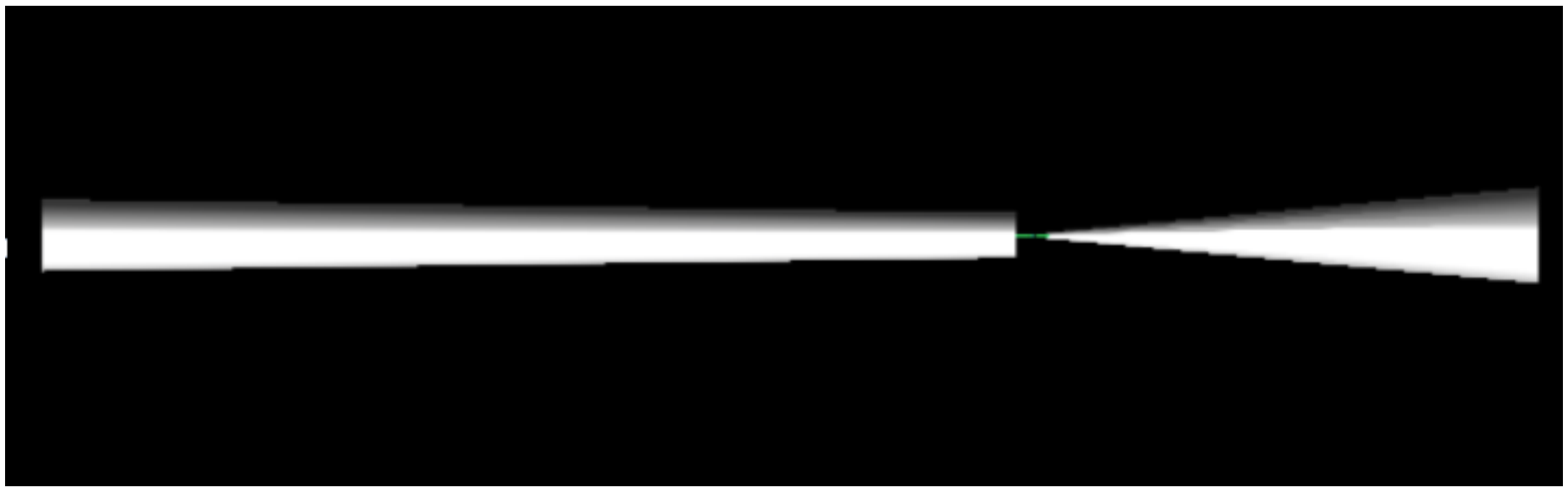

**Figure 5:** A drawing of the highest scoring antenna after 15 generations.

### 3.1 Bicone Antennas with Nonlinear Sides

The GENETIS project is currently working on allowing the GA to evolve bicone antennas that have nonlinear sides. By increasing the parameter space by allowing nonlinear sides, it may be possible to further optimize the design and increase the fitness score of the antennas. With this improvement, additional genes are the coefficients of a polynomial that describes the shape of the side. Thus the prior gene of the opening angle is superseded in the newer design. The minor radius and the length are still included in the evolution and the bicones are still vertically asymmetric with different polynomials for the top and bottom cones.

## 4. Antenna Response Evolution Algorithm (AREA)

AREA is an algorithm developed for the evolution of RF antenna responses for detecting UHE neutrinos. By evolving antenna responses instead of physical antennas, AREA can provide insight into the optimal antenna design in the absence of constraints, while also decreasing computation time by removing the need to simulate antenna responses. Optimizing the antenna response provides insight into the behavior of an optimal antenna that may be evolved by PAEA–the optimal antenna should have a response similar to the optimized pattern from AREA.

AREA uses a linear sum of 13 azimuthally symmetric, spherical harmonic functions to model the gain or phase pattern of an antenna. Since we are assuming azimuthal symmetry, we only consider spherical harmonics with the magnetic quantum number $m = 0$. This allows the convoluted form of a gain or phase pattern to be described with the 13 coefficients $a_\ell$ (the orbital angular momentum quantum number) of the spherical harmonics. AREA follows the normal GA procedure with the proportions of the selection methods and genetic operators shown in Table 2.

**Table 2:** Summary of the selection methods and operators used in the AREA procedure

| Fraction | Selection Method | Operator |
|---|---|---|
| 1/2 | Roulette | Crossover |
| 1/6 | Tournament | Crossover |
| 1/6 | Roulette | Mutation |
| 1/6 | Tournament | Mutation |

The fitness score for each individual is evaluated based on simulated neutrino detection rates for various azimuthal angles. AREA uses a simplified version of AraSim, called AraSim Lite.AraSim Lite simplifies AraSim's fitness function by omitting the simulation's ray-tracing, noise waveforms, signal polarization, and ice modeling, thereby reducing the computational run time[15]. As with PAEA, AREA uses the effective ice volumeas the fitness score.

The results of an initial AREA test are presented in Figure 6. On the left, the change in the maximum and average fitness scores of a population are given over 500 generations, showing a convergence in less than 100 generations with minimal improvement thereafter. On the right, a final evolved radiation pattern is shown. Results indicate that antennas that are more sensitive to up-going signals are more desirable, which matches the current results from PAEA. Ongoing improvements to AREA include utilizing the complete AraSim instead of AraSim Lite, allowing for more accurate fitness scores for the evolved antenna responses in future runs.

## 5. Conclusions

These proceedings provide an overview of and update on the optimization projects currently being undertaken by the GENETIS collaboration. GAs are an efficient machine learning method for maximizing the fitness of the generated individuals. PAEA uses the optimized GA to evolve bicone antenna designs, using six independent genes for each antenna. The fitness score is determined through AraSim and compared to the antennas currently in use in the ARA experiment. So far, 15 generations have been evolved, with the average and maximum fitness score of the evolution now exceeding the performance of the ARA bicone. AREA is being used to optimize beam patterns for antennas to be used in UHE neutrino radio detection experiments. Evolving for beam patterns allows for greater complexity and removes the need to simulate antenna responses, reducing computation time, but doesn't result in an antenna geometry. This allows for a better understanding of the optimized gain pattern without any concern for antenna design constraints. Future runs utilizing the complete AraSim will allow more accurate fitness score calculations.

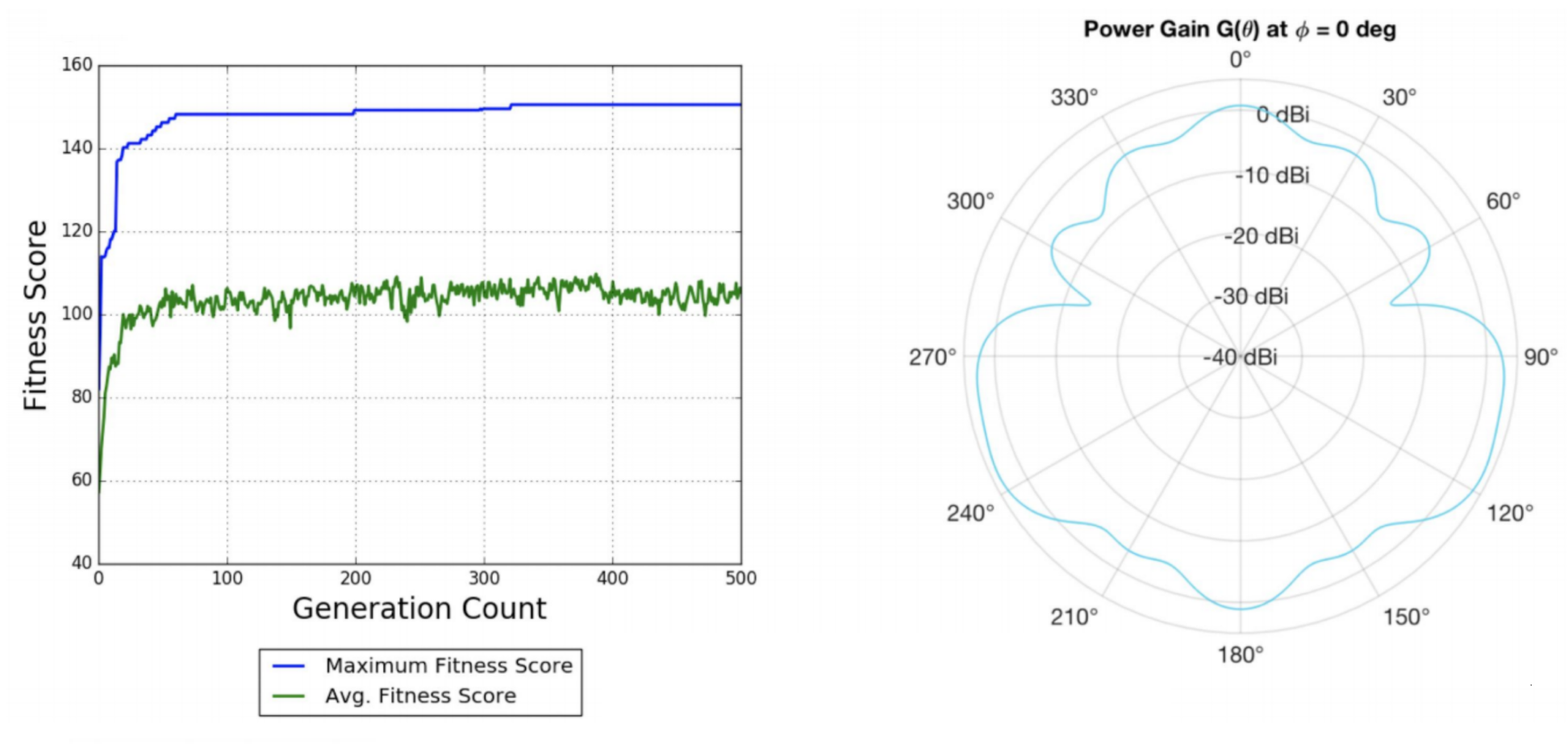

**Figure 6:** Results of an example AREA procedure for energy $10^{18}$ eV: (a) The change in maximum and average fitness score over 500 generations; (b) The radiation pattern with the best fitness score after 500 generations [15].

Further research will involve increasing the parameter space being explored by PAEA and further improving the performance and efficiency of the GA. An improvement to the fitness calculations will reduce computation time by more efficiently using AraSim to separately simulate the neutrino generation and the detector performance. Additionally, new techniques are being tested in the GA to improve the performance. Future research will seek to apply similar techniques to other aspects of astrophysics experiments, such as different antenna designs, array geometries, and trigger systems.

## Acknowledgements

The GENETIS team is grateful for support from the Ohio State Department of Physics Summer Undergraduate Research program, support from the Center for Cosmology and Astroparticle Physics, and the Cal Poly Connect Grant. We would also like to thank the Ohio Supercomputing Center. J. Rolla would like to thank the National Science Foundation for support under Award 1806923 and the Ohio State University Alumni Grants for Graduate Research and Scholarship.

## Full Authors List: GENETIS Collaboration

Julie Rolla[1], Dean Arakaki[2], Maximilian Clowdus[1], Amy Connolly[1], Ryan Debolt[1], Leo Deer[1], Ethan Fahimi[1], Eliot Ferstl[1], Suren Gourapura[1], Corey Harris[2], Luke Letwin [2], Alex Machtay[1], Alex Patton[1], Carl Pfendner[3] Cade Sbrocco[1], Tom Sinha[1], Ben Sipe[1], Kai Staats[4] Jacob Trevithick[2], and Stephanie Wissel[5]

[1]Ohio State University [2]California Polytechnic State University [3] Denison University [4]Arizona State University [5]Pennsylvania State University